\documentclass[a4paper,11pt]{article}
\usepackage{jheppub}
\usepackage{graphicx,amsthm}

\newcommand{\dd}[1][]{\mathrm{d}^{#1}}
\newcommand{\imag}{i}
\theoremstyle{remark}

\newcommand{\Graph}[2][0.3]{\vcenter{\hbox{\includegraphics[scale=#1]{#2}}}}

\makeatletter
\renewcommand\@fpheader{}
\renewcommand\@journal{}
\makeatother

\title{A quasi-finite basis for multi-loop Feynman integrals}

\preprint{MITP/14-076}

\author[a]{Andreas von Manteuffel,}
\author[b,\,c]{Erik Panzer,}
\author[\,a]{and Robert M. Schabinger}

\affiliation[a]{
PRISMA Cluster of Excellence \& Institute of Physics\\
Johannes Gutenberg University, 55099 Mainz, Germany}
\affiliation[b]{
Institutes of Physics and Mathematics,
Humboldt-Universit\"{a}t zu Berlin,\\
Unter den Linden 6, 10099 Berlin, Germany}
\affiliation[c]{
Institute des Hautes \'{E}tudes Scientifiques, Le Bois Marie,\\
35 Route de Chartres, 91440 Bures-sur-Yvette, France}

\abstract{
We present a new method for the decomposition of multi-loop Euclidean Feynman integrals into
quasi-finite Feynman integrals.
These are defined in shifted dimensions with higher powers of the propagators,
make explicit both infrared and ultraviolet divergences,
and allow for an immediate and trivial expansion in the parameter of dimensional regularization.
Our approach avoids the introduction of spurious structures and thereby leaves integrals particularly
accessible to direct analytical integration techniques.
Alternatively, the resulting convergent Feynman parameter integrals may be evaluated numerically.
Our approach is guided by previous work by the second author but overcomes
practical limitations of the original procedure by employing integration by parts reduction.
}

\emailAdd{manteuffel@uni-mainz.de}
\emailAdd{erikpanzer@ihes.fr}
\emailAdd{rschabin@uni-mainz.de}

\keywords{%
QCD,
Feynman integrals,
NLO and NNLO Calculations}

\begin{document}
\maketitle
\section{Introduction}
\label{sec:intro}
Since the early days of quantum field theory, infrared and ultraviolet divergences in the perturbative expansions of $n$-point correlation functions have proven
to be a technical obstruction which has hindered our ability to make precise predictions about the subatomic world. Although the development of dimensional regularization~\cite{'tHooft:1972fi} and the realization that it allowed
for an elegant and unified treatment of both infrared and ultraviolet divergences~\cite{Marciano:1974tv} were major milestones, decades elapsed before a scheme was proposed for the resolution of divergences
in general Euclidean Feynman integrals~\cite{Binoth:2000ps}.
This method, commonly referred to as sector decomposition, has subsequently been refined to guarantee that the procedure eventually terminates~\cite{Bogner:2007cr}
and to allow for numerical evaluations in physical kinematics~\cite{Borowka:2012yc}.
In recent years, applications of the sector decomposition technique have become commonplace and
various flavors of it have been implemented in widely-used public software packages~\cite{Bogner:2007cr,Borowka:2012yc,Carter:2010hi,Borowka:2013cma,Smirnov:2008py,Smirnov:2009pb,Smirnov:2013eza}.

Although sector decomposition allows for a numerical treatment of rather complicated multi-scale Feynman integrals~\cite{Borowka:2013cma},
the method typically produces a large number of convergent parametric integrals which may be difficult to treat analytically
due to the fact that they have no obvious graph-theoretic interpretation. 
In a sector decomposition-based approach, the relevant Feynman parameter integrals are split up in order to extract the poles in $\epsilon$,
the parameter of dimensional regularization, via subtraction terms.
This procedure involves reparametrizations and modifies the structure of the integrands substantially, to the extent that it is not straightforward to apply modern
analytical integration strategies which are known to work for convergent Feynman integrals.

In general, a decomposition of the integration domain can lead to spurious structures
in the individual integrals which, however, cancel in the sum.
For instance, suppose that the $\epsilon$ expansion of the integral (inspired by the discussion in reference \cite{Anastasiou:2003gr})
\begin{equation}
I(\epsilon) = \int_0^1 {\rm d}t ~t^{-1-\epsilon}(1-t)^{-1-2\epsilon} {}_2F_1(\epsilon,1-\epsilon;-\epsilon;t)
\end{equation}
must be calculated to $\mathcal{O}\left(\epsilon\right)$. A direct evaluation of the integral is possible in this case,
\begin{equation}
I(\epsilon) = -\frac{(1+2\epsilon)\Gamma(-\epsilon)\Gamma(-1-3\epsilon)}{\Gamma(-4\epsilon)}\,,
\end{equation}
and it follows that the desired result is
\begin{equation}
\label{eq:I}
I(\epsilon) = -\frac{4}{3\epsilon}+\frac{4}{3}+ \left(-4 + \frac{2}{3}\pi^2\right)\epsilon + \mathcal{O}\left(\epsilon^2\right)\,.
\end{equation}

Alternatively, one could attempt to carry out a sector decomposition of $I(\epsilon)$ analytically. 
A natural but essentially arbitrary choice would be to split the integral up at the point $t = 1/2$ and write $I(\epsilon) = I_1(\epsilon) + I_2(\epsilon)$, where
\begin{align}
I_1(\epsilon) &= \int_0^{1/2} {\rm d}t ~t^{-1-\epsilon}(1-t)^{-1-2\epsilon} {}_2F_1(\epsilon,1-\epsilon;-\epsilon;t)\quad \text{and}\\
I_2(\epsilon) &= \int_{1/2}^1 {\rm d}t ~t^{-1-\epsilon}(1-t)^{-1-2\epsilon} {}_2F_1(\epsilon,1-\epsilon;-\epsilon;t)\,.
\end{align}
At this point, both $I_1(\epsilon)$ and $I_2(\epsilon)$ can be calculated by remapping their integration domains to the unit interval, 
expanding them in plus distributions under the integral sign to the required order in $\epsilon$,
and then, finally, evaluating the resulting finite integrals analytically. Going through these steps, we find
\begin{align}
\label{eq:I1}
I_1(\epsilon) &= -\frac{1}{\epsilon}-1+\left(3 + \frac{1}{3}\pi^2-8\ln(2)\right)\epsilon + \mathcal{O}\left(\epsilon^2\right)\quad \text{and}\\
\label{eq:I2}
I_2(\epsilon) &= -\frac{1}{3\epsilon}+\frac{7}{3}+\left(-7 + \frac{1}{3}\pi^2+8\ln(2)\right)\epsilon + \mathcal{O}\left(\epsilon^2\right)\,.
\end{align}

Clearly, Eqs. (\ref{eq:I1}) and (\ref{eq:I2}) are more complicated than Eq. (\ref{eq:I}); the point is that, by rewriting $I(\epsilon)$ as the sum of $I_1(\epsilon)$ and $I_2(\epsilon)$,
we are forced to speak about spurious transcendental numbers such as $\ln(2)$ which do not appear in the actual result of interest. When one applies the sector decomposition technique to more complicated classes of parametric integrals,
a more serious problem arises: spurious transcendental functions\footnote{That is to say, transcendental functions which do not appear in the final result when it is written in an appropriate normal form free of hidden zeros.}
may be encountered at intermediate stages of the calculation as well~\cite{vonManteuffel:2013vja}. In favorable situations, one may be able to treat the new functions by considering a suitable generalization of the function
space. However, to the best of our knowledge, there is no systematic way to do this in situations more complicated than the one treated in~\cite{vonManteuffel:2013vja} and, therefore, the problem simply cannot be ignored.

Earlier this year, an alternative systematic strategy for the extraction of divergences in general multi-loop Euclidean Feynman integrals was presented by the second author~\cite{Panzer:2014gra}.
Starting from the Feynman parametric representation of a given integral, the method applies an appropriate sequence of partial integrations which ultimately render all Feynman parameter integrations which must be carried out 
finite in the limit $\epsilon\to 0$. In what follows, we will often refer to these particular partial integrations as {\it regularizing dimension shifts}. 
As we discuss in detail in Section~\ref{sec:method}, this regularization method expresses the original Feynman integral in terms of so-called {\it quasi-finite} Feynman integrals,
Feynman integrals which have, at worst, a $1/\epsilon$ pole which originates from the Gamma function prefactor in the Feynman parameter representation of the integral.
In particular, quasi-finite integrals are free of what we call {\it subdivergences}, divergences which arise from the Feynman parameter integrations themselves.
As we will discuss in detail below, these quasi-finite integrals resemble the original Feynman integrals (they are built out of the same set of propagators)
but they may live in shifted dimensions and some propagators may enter raised to higher powers (dots).

This kind of singularity resolution allows one to perform the $\epsilon$ expansion at the level of the integrand without employing subtraction terms
and it thereby expresses each coefficient of the Laurent series in terms of well-behaved, convergent integrals.
The ability to write the Laurent expansion of a given Feynman integral in such a form has important implications:
\begin{itemize}
  \item \emph{Numerical evaluation.} 
  The representation allows for an immediate numerical evaluation in the Euclidean region, at least in principle, without further
  preparations.\footnote{Note that the integrands may still have integrable logarithmic endpoint singularities which, in a numerical evaluation, could have a negative impact on the convergence and thus require appropriate treatment.}
  \item \emph{Analytical evaluation.}
  All resulting parametric integrals are themselves Feynman integrals and are therefore well-adapted to direct analytical integration techniques.
\end{itemize}

Let us motivate in particular the second point, which was the original trigger for the development of the method
and is what distinguishes it from the conventional sector decomposition approach.
In brief, {\tt HyperInt}~\cite{Panzer:2014caa} is a computer algebra package for the analytical evaluation of parametric integrals which can be expressed in terms of multiple polylogarithms and,
in particular, Feynman integrals which happen to be {\it linearly reducible}. 
Roughly speaking, a Feynman integral is linearly reducible if some ordering of the Feynman (or Schwinger) parameters in its parametric representation exists which has the property that one can
integrate out all variables in this particular order without encountering functions which cannot be expressed in terms of multiple polylogarithms. 
Although this criterion may at first sight seem too restrictive to be of any use, it turns out that, in Euclidean kinematics, it allows for a highly-automated, analytical evaluation of numerous
single- and multi-scale, multi-loop Feynman integrals of phenomenological and mathematical interest~\cite{Panzer:2014gra,Panzer:2014caa,Panzer:2013cha}.\footnote{%
The integration via hyperlogarithms implemented in {\tt HyperInt} was first proposed in \cite{Brown:2008um} and the method by which the program decides whether a given Feynman integral is in fact
linearly reducible is based on the compatibility graph method introduced in \cite{Brown:PeriodsFeynmanIntegrals}. See references~\cite{Panzer:2014caa} and~\cite{Panzer:PhD} for further details.} 
It seems to offer an attractive method for the evaluation of Feynman integrals, especially for those for which the method of differential equations does not provide a complete solution.\footnote{It is worth pointing out 
that sometimes even non-trivial, multi-scale Feynman integrals cannot be completely determined from the differential equations that they satisfy and associated regularity conditions.}
The method of direct integration employed by {\tt HyperInt} makes sense only for well-defined, convergent integrals and it is therefore essential that all subdivergences be resolved beforehand by passing to quasi-finite integrals.
Conventional sector decomposition is in general of little use here because it involves changes of variables which can destroy the linear reducibility.

In fact, an elementary implementation of this singularity resolution method is already part of the {\tt HyperInt} package, but, for some integrals, it fails to produce useful results due to runaway expression swell.
For example, the well-studied~\cite{1001.2887,1004.3653,1010.1334}, nine-line, planar master integral for the three-loop form factor in massless QCD ($A_{9,1}$ in the notation of reference~\cite{1004.3653})
will quickly consume more than 30 gigabytes of random-access memory if one attempts a singularity resolution of the integral along the lines suggested by the {\tt HyperInt} implementation. 
The typical situation seems to be that relatively compact and simple linear combinations of quasi-finite integrals are generated by the existing implementation except for {\it top-level topologies} or, in other words,
Feynman integrals whose graphs have the maximal number of lines possible for a fixed number of loops and legs (3-regular graphs).
Furthermore, the method generates spurious poles in $\epsilon$ by introducing linear combinations of Feynman integrals which add to zero.
Clearly, for the {\tt HyperInt} approach to be competitive with other methods in all cases of practical interest, improvement is required.

In this paper, we explain that these problems originate from an arbitrariness within the method of
regularizing dimension shifts which can be resolved by making use of integration by parts (IBP) identities~\cite{Tkachov:1981wb,Chetyrkin:1981qh,Laporta:2001dd}.
We show how IBP reductions, as implemented, for example, in the recent public codes~\cite{vonManteuffel:2012np,Lee:LiteRed,Smirnov:2014hma},
allow for the direct construction of quasi-finite bases in an efficient manner which completely supersedes and replaces the procedure discussed above. 
We describe a new algorithm and demonstrate its advantages by considering a number of real-life examples.
For instance, the two-loop non-planar form factor integral has a Laurent expansion which begins at $\mathcal{O}\left(\epsilon^{-4}\right)$ in $d=4-2\epsilon$ spacetime dimensions
but it can be rewritten in terms of quasi-finite master integrals as (details will be given in Section~\ref{sec:example})
\begin{equation}
\label{eq:formfactor-quasi-finite}%
\begin{split}
	\Graph{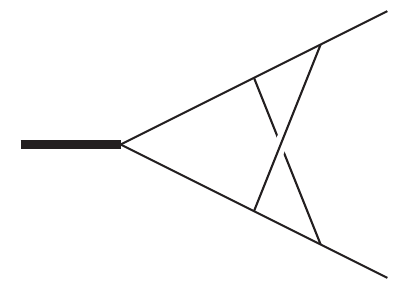}^{(4-2\epsilon)}
	&= \frac{4(1-\epsilon)(3-4\epsilon)(1-4\epsilon)}{\epsilon s^2}~ \Graph{2loopnp}^{(6-2\epsilon)}
	\\ &\quad
	- \frac{10-65\epsilon+131\epsilon^2-74\epsilon^3}{\epsilon^3 s^2} ~\Graph{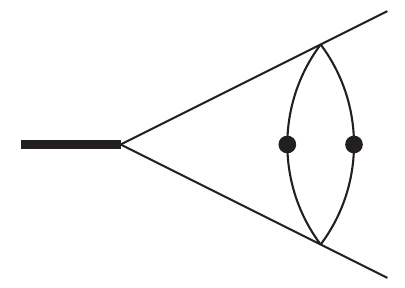}^{(6-2\epsilon)}
	\\ &\quad
	- \frac{14-119\epsilon+355\epsilon^2-420\epsilon^3+172\epsilon^4}{(1-2\epsilon)\epsilon^3 s^3} ~\Graph{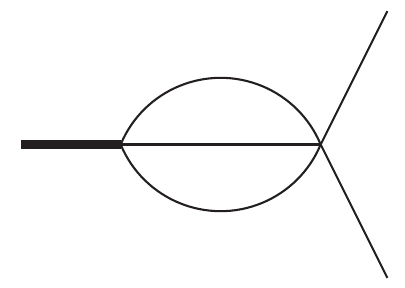}^{(4-2\epsilon)}\,.
\end{split}
\end{equation}
Motivated by the character of the decomposition, we will refer to our singularity resolution procedure as the \emph{minimal dimension shifts and dots method}.

To be clear, our method does not turn a computationally complex problem into a computationally simple one.
Rather, it maps the computationally intensive part of the resolution procedure to a problem which can be solved using integration by parts reduction. 
Of course, in the standard approach, integration by parts reductions are quite expensive and time-consuming to compute.
However, one should remember that multi-loop calculations are often performed using rather challenging IBP reductions anyway 
in order to write the amplitude or interference terms of physical interest as a linear combination of master integrals.
Therefore, the reduction problem may be assumed to be under control provided that quasi-finite integrals exist which look no more daunting to reduce
than the most complicated integrals which appear in the physics result.
This assumption appears to be valid in the examples that we have studied so far and the approach advocated in this work therefore seems quite reasonable to us. 
In any case, excellent prospects exist for the construction of new integration by parts reduction programs which do not suffer from known computational issues which plague traditional solvers~\cite{vonManteuffel:2014ixa};
on general grounds, such programs are expected to perform much better than those which are currently available.
It is worth pointing out that, throughout this article, we assume some familiarity with the standard parlance of the integration by parts reduction field. 
The relevant definitions are given in many places, {\it e.g.} reference~\cite{Studerus:2009ye}.

Finally, let us comment on the scope of our method.
For the most part, we will focus on the important but comparatively simple case of Euclidean kinematics.
However, as will be explained in more detail below, we expect much of our discussion to carry over to
multi-loop Feynman integrals in physical kinematics with minor modifications for many cases of practical interest.
At the present time, completely general physical Feynman integrals are not covered by our method.
A systematic treatment of completely general multi-loop Feynman integrals could benefit from techniques developed for the method of regions~\cite{Beneke:1997zp,Smirnov:1998vk,Smirnov:1999bza,Pak:2010pt,Jantzen:2012mw},
but such an analysis is well beyond the scope of this work.
Needless to say, there are still other interesting classes of integrals, {\it e.g.} phase space integrals, which we do not discuss in our study.

In summary, we propose a new method which uses standard IBP reductions to express Euclidean multi-loop integrals in terms of an alternative, quasi-finite basis of integrals which is characterized by a complete absence of subdivergences.
As its name suggests, the minimal dimension shifts and dots method is a compact and efficient procedure which completely supersedes the method of regularizing dimension shifts implemented in {\tt HyperInt}.
Our technique makes the singularities of Euclidean multi-loop Feynman integrals explicit and produces results which are well-suited for subsequent analytical and/or numerical evaluations.
In Section~\ref{sec:method}, after reviewing the method of regularizing dimension shifts and its shortcomings, we 
demonstrate the existence of quasi-finite integral bases and then explain how to carry out the decomposition of an arbitrary integral with our new method. We also discuss how, with minor modifications, it may be possible to extend 
our Euclidean space method to treat physical multi-loop Feynman integrals. 
As explicit examples, we present the minimal dimension shifts and dots singularity resolution of the two-loop non-planar form factor integral and
a minimal quasi-finite basis for the two-loop planar double box integral family in Section~\ref{sec:example}.
Finally, in Section \ref{sec:conclusions}, we conclude and discuss some interesting directions for future research.

\section{The Method}%
\label{sec:method}

\subsection{Regularizing dimension shifts}
\label{sec:DDshifts}
\begin{figure}
	\centering%
	\includegraphics[scale=0.45]{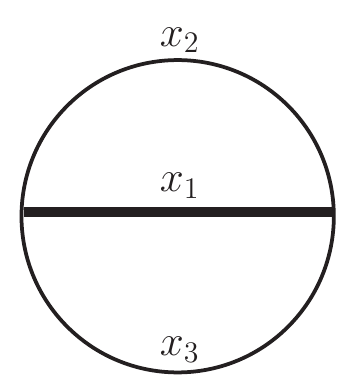}%
	\caption{The two-loop tadpole diagram with a single massive propagator (thick line).}%
	\label{fig:tadpole}%
\end{figure}

We begin with a simple example to remind the reader how the method of regularizing dimension shifts works.
Consider the two-loop tadpole diagram with a single massive line depicted in Figure~\ref{fig:tadpole}.
The corresponding scalar Feynman integral, hereafter referred to as $T_{111}(m^2, 4 - 2\epsilon)$, works well as a test case because the result in $d=4-2\epsilon$ is available in closed form,
\begin{align}
	T_{111}(m^2, 4 - 2\epsilon) 
	&= \int\!\frac{\dd[d] k_1}{\imag \pi^{d/2}} \int\!\frac{\dd[d] k_2}{\imag \pi^{d/2}} 
			\frac{1}{\left( (k_1+k_2)^2 - m^2\right) k_1^2 k_2^2}
	\nonumber \\
	&= -\left(m^2\right)^{1-2\epsilon}
		\frac{
					\Gamma(-1 + 2\epsilon)\Gamma(\epsilon)\Gamma(1-\epsilon)
		}{
					1-\epsilon
		}\,,
	\label{eq:exclosed}%
\end{align}
and the integral is inherently Euclidean in nature.\footnote{For the sake of definiteness, we work in Minkowski spacetime with a $+i 0$ pole prescription.}
Its parametric representation reads
\begin{equation}
	T_{111}(m^2, 4 - 2\epsilon) 
	= -\Gamma(-1+2\epsilon)
		\int_0^\infty\!\!\!\!  \dd x_1 \delta(1-x_1) \int_0^\infty\!\!\!\!  \dd x_2 \int_0^\infty\!\!\!\!  \dd x_3
		\ 
		\mathcal{U}^{-3+3\epsilon}\mathcal{F}^{1-2\epsilon}\,,
	\label{eq:normalparamrep}%
\end{equation}
where $\mathcal{U}$ and $\mathcal{F}$ are respectively the first and second Symanzik polynomials \cite{BognerWeinzierl:GraphPolynomials},
\begin{equation}
	\mathcal{U} 
	= x_1 x_2 + x_1 x_3 + x_2 x_3
	\quad\text{and}\quad
	\mathcal{F} 
	= m^2 x_1 \mathcal{U}\,.
\end{equation}
Eq. \eqref{eq:normalparamrep} cannot be expanded in $\epsilon$ as-is because we see from Eq. \eqref{eq:exclosed} that
$T_{111}(m^2, 4 - 2\epsilon)$ has an $\epsilon$ expansion which begins at $\mathcal{O}\left(\epsilon^{-2}\right)$ and, therefore, a subdivergence.

The first step of the resolution procedure is to identify those components of the integration domain which
lead to a divergence of the integral if the parameter $\epsilon$ of dimensional regularization is set to zero.
For Euclidean integrals, such an analysis is particularly simple. In Euclidean kinematics, all monomials in $\mathcal{U}$ and $\mathcal{F}$ are positive in the interior of the integration domain
and, therefore, all possible divergences can be detected by studying the behavior of the integrand at the integration boundaries.
Moreover, it turns out that a particularly simple power-counting analysis is sufficient~\cite{Panzer:2014gra} to identify and subsequently extract the divergences in this case:
one inspects, for all non-empty proper subsets of Feynman parameters, the behavior of the integrand as these parameters or their reciprocals are uniformly taken to approach zero.
While this simple recipe effectively covers many non-Euclidean Feynman integrals of phenomenological interest by virtue of the principle of analytical continuation,
we emphasize that fully general Feynman integrals may have {\it e.g.} additional non-integrable divergences in the interior and in general require a more elaborate power counting analysis.
In the following, we restrict ourselves to Euclidean integrals and further simplify our notation by considering only the case where some Feynman parameters
approach zero (for many examples, including those that we consider below, this turns out to be sufficient).

For $T_{111}(m^2, 4 - 2\epsilon)$ we therefore consider the six subsets\footnote{The presence of the delta function constraint in Eq.~\eqref{eq:normalparamrep} allows us to discard the improper subset $\{x_1,x_2,x_3\}$ at the outset.}
\begin{align*}
	&\{x_1,x_2\}, \qquad\qquad \{x_1,x_3\}, \qquad\qquad \{x_2,x_3\},\nonumber \\
	&~~\{x_1\}, \qquad\qquad~~~~\, \{x_2\}, \qquad\qquad~~~~\, \{x_3\}.
\end{align*}
As explained in~\cite{Panzer:2014gra}, power counting associates to each such subset $J$ an index which characterizes the relevance of the subset.
In what follows, we will call a parameter subset which is associated with
a subdivergence a \emph{relevant subset} and one which is not an \emph{irrelevant subset}.
To define this \emph{degree of divergence}, $\omega_J(P)$, we must first define the asymptotic degree of a parameter subset $J$ with respect to the parametric integrand $P$ of the Feynman integral.
Let $P_{J_\lambda}$ denote the integrand $P$ but with the Feynman parameters in $J$ replaced by $\lambda J$.
For example, if we take 
\begin{equation}
P =
\mathcal{U}^{-3+3\epsilon}\mathcal{F}^{1-2\epsilon}
\end{equation}
from Eq. \eqref{eq:normalparamrep} and set $J = \{x_2, x_3\}$, then
$\lambda J = \{\lambda x_2, \lambda x_3\}$ and we find
\begin{equation}
	P_{J_{\lambda}}
	=
\lambda^{\epsilon-2}\left(m^2 x_1 \right)^{1-2\epsilon} \left( x_1 x_2 + x_1 x_3 + \lambda x_2 x_3 \right)^{\epsilon-2}\,.
	\label{eq:tadpole-rescaled-integrand}%
\end{equation}
The asymptotic degree of $J$ with respect to $P$, ${\rm deg}_J(P)$, is simply the unique number $s$ such that
\begin{equation}
	\lim_{\lambda \to 0} \left\{\lambda^{-s} P_{J_\lambda}\right\}
\end{equation}
exists and is non-zero. Finally, if we let $|J|$ denote the number of elements of $J$, we have
\begin{equation}
	\omega_J(P) 
	= |J| + {\rm deg}_J(P)\,,
\end{equation}
where, as before, $P$ is the parametric integrand under consideration.

Once all indices have been computed, the results allow for the determination of an upper bound on the number of singularities which must be resolved. 
In fact, the only subsets of interest are those which satisfy
$
	\displaystyle \lim_{\epsilon \to 0} \left\{\omega_J(P) \right\}
	\leq 0
$; all other parameter subsets are irrelevant and can safely be discarded.
For the example at hand, only $J=\{x_2, x_3\}$ survives this culling process and from Eq. \eqref{eq:tadpole-rescaled-integrand} we read off $\deg_J(P) = \epsilon-2$ and finally obtain
\begin{equation}
	\omega_{\{x_2, x_3\}}\left(P\right) = \epsilon \,. 
\end{equation}
In general, each surviving (\emph{relevant}) subset $J$ describes a non-integrable singularity of the parametric integrand $P$
which can subsequently be resolved by carrying out some number of partial integrations determined by the index $\omega_J(P)$.

The singularity of $P$ associated to $J$ can be resolved by a sequence of integrand transformations
such that the result corresponds again to a linear combination of Feynman integrands~\cite{Panzer:2014gra}.
Each of these transformations proceeds in the following way.
We insert a factor
\begin{equation}
1 = \int_0^\infty\!\!\!\! \mathrm{d}\lambda\,\, \delta(\lambda - x_J)
\end{equation}
with $x_J = \sum_{j\in J} x_j$, rescale $x_j\to \lambda x_j$ for all $j\in J$,
suitably account for the delta functions,
and perform a partial integration according to
\begin{equation}
\int_0^\infty\!\!\!\! \mathrm{d}\lambda\,\,  \lambda^{|J|-1} P_{J_\lambda} =
 \left. \frac{1}{\omega_J(P)} \lambda^{|J|}  P_{J_\lambda} \right|_{\lambda=0}^{\infty}
  - \frac{1}{\omega_J(P)} \int_0^\infty\!\!\!\! \mathrm{d}\lambda\,\,  \lambda^{\omega_J(P)}
    \frac{\partial}{\partial\lambda} \left(\lambda^{-{\rm deg}_J(P)} P_{J_{\lambda}}\right)
\end{equation}
where the surface term vanishes.
From these considerations, we obtain the new integrand
\begin{equation}
	\label{eq:dimregpartial}
	P^\prime = - \frac{1}{\omega_J(P)} \frac{\partial}{\partial \lambda}\left(\lambda^{-{\rm deg}_J(P)} P_{J_{\lambda}}\right)\bigg|_{\lambda \to 1}\,.
\end{equation}
The singularity structure of the transformed integrand, $P^\prime$, is better than that of the original integrand by design.
However, it should be emphasized that, if $\displaystyle \lim_{\epsilon \to 0} \left\{\omega_J(P) \right\} < 0$, additional work is likely required because the parameter subset $J$
will most probably still have a relevant index (now with respect to the transformed integrand) of the form
$\displaystyle \lim_{\epsilon \to 0} \left\{\omega_J\left(P^\prime\right) \right\} = 1 + \lim_{\epsilon \to 0} \left\{\omega_J(P) \right\} \leq 0$ after the first partial integration.
This means that as many as 
$
	\displaystyle 1 + |\lim_{\epsilon \to 0} \left\{\omega_J(P) \right\}|
$
partial integrations may be required to completely resolve the singularity associated with the parameter subset $J$.\footnote{%
Only the last of these partial integrations introduces a denominator $\omega_J(P)$ into \eqref{eq:dimregpartial} which vanishes as $\epsilon\rightarrow 0$, so every subdivergence
(whether logarithmic or worse) contributes at most a simple pole in $\epsilon$.}

Returning to our treatment of $T_{111}(m^2, 4 - 2\epsilon)$, we arrive at
\begin{align}
	T_{111}(m^2, 4 - 2\epsilon)
	&= -\frac{2-\epsilon}{\epsilon}\,\Gamma(-1+2\epsilon) \int_0^\infty\!\!\!\!  \dd x_1 \delta(1-x_1) \int_0^\infty\!\!\!\! \dd x_2 \int_0^\infty\!\!\!\!  \dd x_3 \,
	 x_2 x_3 \left(m^2 x_1\right)^{1-2\epsilon}\times
	\nonumber\\
	&\quad\times\left(x_1 x_2 + x_1 x_3 + x_2 x_3\right)^{\epsilon-3}
	\nonumber\\
	&=  -\frac{2-\epsilon}{\epsilon} T_{122}(m^2, 6 - 2\epsilon)
	\label{eq:normalfinalrep}%
\end{align}
by taking Eq. \eqref{eq:tadpole-rescaled-integrand}, applying Eq.~\eqref{eq:dimregpartial}, and then reinterpreting the resulting parametric integral as $T_{122}(m^2, 6 - 2\epsilon)$.
Here, the subscripts indicate the presence of squared propagators on edges $2$ and $3$ of the graph (see Figure \ref{fig:tadpole}).
By expanding this quasi-finite representation of $T_{111}(m^2, 4 - 2\epsilon)$ in $\epsilon$ using {\it e.g.} the function {\tt Series} provided by the {\tt Mathematica} computer algebra system and then integrating term-by-term\footnote{%
In this simple case, any order of the variables is linearly reducible and the integrations are straightforward to perform analytically using either {\tt Maple} or {\tt Mathematica} out of the box.},
one can easily check that the result is in complete agreement with the expansion of the exact result, Eq.~\eqref{eq:exclosed}.

\subsection{Problems with the transformations}
\label{sec:problems}
As a matter of principle, the method of regularizing dimension shifts applies to arbitrary Euclidean Feynman integrals. However, while the elementary version of the method works well for simple examples such as $T_{111}(m^2, 4 - 2\epsilon)$
from the previous section, it effectively fails in many non-trivial situations of practical interest because it produces expressions which are orders of magnitude too large to be processed further.
In fact, it has several undesirable features:
\begin{itemize}
	\item \emph{Proliferation of terms.} 
		Each partial integration carried out according to the prescription provided by Eq. \eqref{eq:dimregpartial} can produce a large number of terms,
		as we will demonstrate explicitly in Eq. \eqref{eq:dimshift} for the example of the two-loop non-planar form factor integral.
		If the integral under consideration has multiple subdivergences, the recursive resolution of its singularities
		can generate an exponentially growing number of terms due to repeated applications of regularizing dimension shifts.
		In practice this expression swell can render the method essentially useless.
	\item \emph{Ambiguity.}
		When several subdivergences are present, one can resolve them in any order. 
		But, in general, the resolution of a given singularity may cause the behavior of other parameter subsets to improve, 
		possibly to the point where they become irrelevant and can be discarded. The chosen order can thus have a drastic effect on the number of quasi-finite integrals which appear in the final expression for the integral of interest.
		This problem is compounded by the fact that it is not at all clear how to choose a good ordering of the parameter subsets at the outset.

	\item \emph{Spurious singularities.}
		In some cases the method generates spurious higher-order poles in $\epsilon$ multiplying a linear combination of integrals, which in fact is a hidden zero (at least up to a certain order in the $\epsilon$ expansion).
		In particular, it may occur that terms proportional to {\it e.g.} $\epsilon^{-2 L - 1}$ are produced for $L$-loop integrals which are expected to have, at worst, $\epsilon^{-2 L}$ poles
		(for instance, see~\cite[Example~6.1]{Panzer:2014gra}).
                Clearly, it is of great interest to avoid such spurious structures if possible.
\end{itemize}

All of the problems listed above are related and stem from the fact that, generically, there will be unresolved linear relations among the various integrals
which emerge from the regularizing dimension shifts.
Fortunately, this problem can be solved by simply expressing all terms which appear at intermediate stages as linear combinations of a small number of master integrals.

\subsection{Integration by parts reduction}
\label{sec:ibp}

It is well-known that, for a given topology and its subtopologies, there is a finite number of linearly independent integrals~\cite{Smirnov:2010hn,Lee:2013hzt}, which are usually referred to as master integrals.
IBP reduction is the standard method used to reduce a given integral with respect to a set of master integrals.
Our aim is to employ IBP reductions to resolve singularities by rewriting divergent integrals as linear combinations of quasi-finite master integrals.
The proof of concept is based on the regularizing dimension shifts discussed above, which introduces Feynman integrals in shifted dimensions $d, d+2, \ldots, d+2\delta$.
Let us therefore describe how IBP reductions can be used to express some integral $b$ of interest in terms
of a linearly independent and complete set of basis integrals, $B^\prime$, each of which may be defined in a different number of dimensions.

If all involved integrals happen to live in the same dimension,
the problem can be solved immediately using any standard IBP reduction program.
In this work, we made use of both {\tt Reduze\;2} and {\tt LiteRed} to carry out our IBP reductions. 
At the present time, the latter program is particularly convenient to use for our purposes since it already ships with built-in support for dimension shifts.
In any case, it is conceptually straightforward to perform the necessary IBP reductions even when using a program without native support for integrals which inhabit different spacetime dimensions.

Suppose we have an IBP reduction to some basis $B_d$ in hand for a fixed number of dimensions, $d$.
Using a dimension shift relation such as the one given in Eq.\ \eqref{eq:dimregpartial}
or the one introduced by Tarasov \cite{Tarasov:1996br,Lee:2009dh,Lee:2010wea},\footnote{
Dimension-shift relations for specific classes of loop integrals were discussed in earlier work \cite{Derkachov:1990osa,Bern:1992em,Bern:1993kr}.}
we can express the basis $B_d$ in terms of integrals in $d+2$ dimensions.
Employing our reductions with respect to $B_d$ but with $d$ replaced by $d+2$, we obtain the relation $B_d = \underline{M}_{\, d} B_{d+2}$.
If the integral $b$ is defined in $d+2\delta_b$ ($0\leq \delta_b\leq \delta$) dimensions, we first reduce it with respect to $B_{d+2\delta_b}$ such that $b = A^\mathrm{T} B_{\,d+2\delta_b}$.
Subsequently, the dimension shift can be used in an iterative fashion to arrive at $b=\tilde{A}^\mathrm{T} B_d = A^\mathrm{T} \underline{M}_{\,d+2\delta_b-2}^{-1} \cdots  \underline{M}_{\,d}^{-1}B_{d}$.
Similar steps for each of the integrals in $B^\prime$ give $B^\prime =\underline{M}^\prime B_{d}$.
Finally, the last relation can be inverted to obtain the desired transformation $b = \tilde{A}^\mathrm{T} {\underline{M}^\prime}^{-1} B^\prime$.

\subsection{Existence of a quasi-finite basis}
\label{sec:existence}

In this section, we show that it is possible to construct a basis of quasi-finite integrals for Euclidean Feynman integrals
which belong to a given topology and its subtopologies.\footnote{Please see Section \ref{sec:DDshifts} for comments on more general Feynman integrals.}
Suppose we start with some choice of master integrals, $B$, out of which at least one integral $b$ is not quasi-finite.
For the sake of definiteness, we adopt the Feynman integral normalization that we used for the integral $T_{111}(m^2, 4 - 2\epsilon)$ treated in Section \ref{sec:DDshifts}. In this normalization,
the parametric representation of $b$, our general, scalar, $L$-loop Feynman integral with $N$ propagators (raised to integer powers $\nu_i$), is given in $d$ dimensions by
\begin{equation}
	b = \frac{\Gamma\big(\nu - \frac{L d}{2}\big) (-1)^{\nu}}{\prod_{i=1}^N \Gamma(\nu_i)} 
	\Bigg[ \prod_{j=1}^N \int_0^{\infty} \dd x_j \Bigg] 
	\delta(1-x_N)
	\,\mathcal{U}^{\nu - (L+1)d/2} \mathcal{F}^{-\nu + L d/2}
	\prod_{k=1}^N x_k^{\nu_k-1}  
	\label{eq:general-parametric-rep}
\end{equation}
and the associated integrand is
\begin{equation}
	P 
	= 
	\mathcal{U}^{\nu - (L+1)d/2} \mathcal{F}^{-\nu + L d/2}
	\prod_{j=1}^N x_j^{\nu_j-1}\,,
	\label{eq:general-parametric-integrand}
\end{equation}
where $\nu = \sum_{i=1}^N \nu_i$ denotes the sum of the multiplicities of the propagators.
If we now perform a regularizing dimension shift with respect to some relevant parameter subset, $J$, using Eq. \eqref{eq:dimregpartial}, the result is
a linear combination of integrands which look like the original but in $d+2$ dimensions and with some number of shifted indices.
As before, we simplify the discussion to divergences at zero and point out that one can treat divergences at infinity or mixed cases in a completely analogous fashion (see~\cite{Panzer:2014gra} for details).
Explicitly, if we factor out the lowest powers of $\lambda$ as
$ \mathcal{U}_{J_{\lambda}} = \lambda^{\deg_{J}(\mathcal{U})} \widetilde{\mathcal{U}} $
and
$ \mathcal{F}_{J_{\lambda}} = \lambda^{\deg_J(\mathcal{F})} \widetilde{\mathcal{F}}$,
then the right-hand side of Eq. \eqref{eq:dimregpartial} becomes
\begin{align}
  P^\prime  &= -\frac{1}{\omega_J(P)}\prod_{j=1}^N x_j^{\nu_j-1} \bigg\{
  \Big(\nu-\tfrac{(L+1)d}{2}\Big) 
  \mathcal{U}^{(\nu + L) - (L+1)(d+2)/2} \mathcal{F}^{-(\nu + L) + L(d+2)/2}
  \frac{\partial \widetilde{\mathcal{U}}}{\partial \lambda}\Big|_{\lambda\rightarrow 1}
  \nonumber\\
  &\quad  - \Big(\nu - \tfrac{L d}{2}\Big) \mathcal{U}^{(\nu + L + 1) - (L+1)(d+2)/2} \mathcal{F}^{-(\nu + L + 1) + L(d+2)/2}
    \frac{\partial \widetilde{\mathcal{F}}}{\partial \lambda}\Big|_{\lambda\rightarrow 1} \bigg\}\,,
  \label{eq:dimregpartial-dimshift}
\end{align}
where
$	\frac{\partial \widetilde{\mathcal{U}}}{\partial \lambda}\big|_{\lambda\rightarrow 1} $
and
$	\frac{\partial \widetilde{\mathcal{F}}}{\partial \lambda}\big|_{\lambda\rightarrow 1} $
are, respectively, polynomials in the $x_i$ of degree $L$ and degree $L + 1$. 
While relations between multi-loop integrals in different spacetime dimensions were written down long ago~\cite{Tarasov:1996br},
the non-trivial and crucial point about this particular shift relation is that {\it all} of the terms it generates have an improved convergence with respect to the parameter subset $J$.

Since this observation is crucial for our basis construction, let us explain it in more detail.
Suppose we consider an arbitrary term of Eq.~\eqref{eq:dimregpartial-dimshift}.
Without loss of generality, one can consider either a numerator monomial of degree $\nu-N+L$
which comes from $\prod_{i=1}^N x_i^{\nu_i-1}\frac{\partial \widetilde{\mathcal{U}}}{\partial \lambda}\big|_{\lambda\rightarrow 1}$ or a numerator monomial
of degree $\nu - N + L + 1$ which comes from $\prod_{i=1}^N x_i^{\nu_i-1}\frac{\partial \widetilde{\mathcal{F}}}{\partial \lambda}\big|_{\lambda\rightarrow 1}$. For the sake of discussion, let us choose a monomial,\\
$m(x_1,\ldots,x_N)\prod_{i=1}^N x_i^{\nu_i-1}$, from $\prod_{i=1}^N x_i^{\nu_i-1}\frac{\partial \widetilde{\mathcal{U}}}{\partial \lambda}\big|_{\lambda\rightarrow 1}$.
 Our chosen Feynman integrand can be written in the suggestive form
\begin{equation}
	-\frac{\nu-(L+1)d/2}{\omega_J(P)} P \frac{m}{\mathcal{U}}
	\label{eq:specific-parametric-rep}
\end{equation}
and, by construction, its degree of divergence with respect to the subset $J$ is better than that of the original integrand.
Specifically, $\deg_J(m) > \deg_J(\mathcal{U})$ since the leading term in $\mathcal{U}_{J_\lambda}$ for $\lambda\to 0$ turns into a $\lambda$-independent term
in $\widetilde{\mathcal{U}}$ which is then annihilated by the action of $\frac{\partial}{\partial \lambda}$ on $\widetilde{\mathcal{U}}$ in Eq. \eqref{eq:dimregpartial-dimshift}.
Therefore, the $\lambda$ power counting of \eqref{eq:specific-parametric-rep} 
with respect to the replacement $J \to \lambda J$ demonstrates that the particular integral $b^\prime$ that we obtained by picking a term of $P^\prime$ at random  is less singular than the integral $b$ that we started with.

Out of several possible choices, we select an integral $b^\prime$ which is linearly independent of $B\setminus b$, which must be possible since, by definition, $B$ is linearly independent.
At this stage, we can replace $b$ by $b^\prime$ in our set of master integrals $B$.
While it may not be the case that $b^\prime$ has fewer poles in $\epsilon$ than $b$, this process
can be repeated until, after a finite number of iterations, we have explicitly constructed a master integral for which the Feynman parameter subset $J$ is irrelevant.
In this fashion, we can proceed subset-by-subset and finally construct a quasi-finite master integral.
We can furthermore repeat the steps described in this section until only quasi-finite master integrals remain in our basis.
Obviously, this algorithm terminates.

\subsection{Finding a minimal quasi-finite basis}
\label{sec:alg}

In the previous section, we showed that, in Euclidean kinematics, it is possible to algorithmically construct a basis of quasi-finite dotted integrals in shifted dimensions.
Such a basis is not unique, and the procedure we described so far has the feature that it may produce master integrals with an unnecessarily large number of dots in highly-shifted spacetime dimensions.
For practical applications, it is usually better to employ an alternative strategy which guarantees an optimal basis choice with respect to some user-defined criterion.
Here, in an effort to keep the IBP reductions as simple as possible, we prefer integrals which have a small number of dots and are defined in a small number of spacetime dimensions.

In fact, one can simply perform a systematic search until one finds enough linearly independent quasi-finite integrals to construct a basis for the topology of interest and all of its subtopologies.
For this, we may consider one topology at a time in a bottom-up approach.
For each topology, we start with the simplest possible integral and then proceed to more complicated integrals by systematically increasing the number of dots and the spacetime dimension.
Our search path is such that all potentially acceptable integral candidates are eventually considered.
For each candidate integral, we check if it is quasi-finite along the lines described in Section \ref{sec:DDshifts}.
In this way, one iterates until enough linearly independent quasi-finite integrals are found to serve as a set of master integrals for each topology.

Having IBP reductions with dimension shifts in hand, one can check the linear independence and completeness of the basis
and, finally, reduce any given integral in the topologies under consideration with respect to the constructed quasi-finite basis.
Using a dedicated implementation, our experiments show that it is possible to generate very large numbers of quasi-finite integrals for a given sector in a short amount of time;
the performance bottleneck is given by the IBP reductions themselves.
We give examples of minimal quasi-finite bases in Section~\ref{sec:example}.

\section{Illustrative Examples}
\label{sec:example}%
\begin{figure}%
	\centering%
  \includegraphics[scale=0.45]{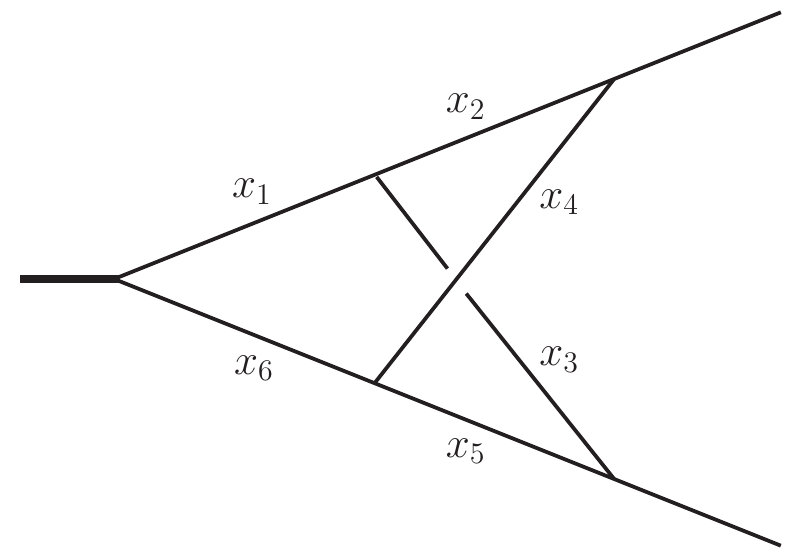}%
  \caption{The two-loop non-planar form factor diagram. The two external momenta on the right are light-like, while the momentum entering from the left squares to $s$.}%
	\label{fig:2loop-crossed-formfactor}%
\end{figure}

In order to clarify how the minimal dimension shifts and dots method proposed in Section~\ref{sec:method} works, let us consider the two-loop non-planar form factor integral depicted in Figure~\ref{fig:2loop-crossed-formfactor}.
This integral, hereafter referred to as $F_{111111}(s, 4 - 2\epsilon)$, turns out to be a particularly clean example.
Its Feynman parameter representation, using the same absolute normalization that we used for $T_{111}(m^2, 4 - 2\epsilon)$ in section \ref{sec:DDshifts}, reads
\begin{equation}
	F_{111111}(s, 4 - 2\epsilon)  = \Gamma(2+2\epsilon) \int_0^\infty \hspace{-1ex}\dd x_1 \cdots \int_0^\infty \hspace{-1ex}\dd x_6 \, \delta(1-x_6) \, \mathcal{U}^{3\epsilon}\mathcal{F}^{-2-2\epsilon}
	\label{eq:formfactor-parametric}%
\end{equation}
with the graph polynomials
\begin{align}
	\mathcal{U} 
	&= (x_1 + x_6)(x_2+x_3+x_4+x_5) + (x_2+x_4)(x_3+x_5)
	\quad\text{and}
	\label{eq:formfactor-U}\\
	\mathcal{F} 
	&= -s(x_1 x_4 x_5 + x_1 x_2 x_6 + x_1 x_3 x_6 + x_2 x_3 x_6 + x_1 x_4 x_6 + x_1 x_5 x_6)\,.
	\label{eq:formfactor-F}%
\end{align}
The subscripts of $F$ denote the powers of each of the six propagators according to the labeling scheme introduced in Figure~\ref{fig:2loop-crossed-formfactor} above.
In particular, a number larger than $1$ corresponds to one or more dots, while a zero corresponds to a pinched (contracted) line.

When we consider this integral at $\epsilon=0$, representation \eqref{eq:formfactor-parametric}
contains many subdivergences.\footnote{Note that the Laurent expansion of $F_{111111}(s, 4 - 2\epsilon)$ in $\epsilon$ begins at $\mathcal{O}\left(\epsilon^{-4}\right)$.}
For example, if we let $P_{111111}(s, 4 - 2\epsilon)$ denote the parametric integrand of $F_{111111}(s, 4 - 2\epsilon)$, then we have
\begin{equation*}
	{\rm deg}_{\{x_1,x_2\}}\left(P_{111111}(s, 4 - 2\epsilon)\right) 
	= -2 -2\epsilon \quad\text{and}\quad
	\omega_{\{x_1,x_2\}}\left(P_{111111}(s, 4 - 2\epsilon)\right) 
	= -2\epsilon
\end{equation*}
and we see that the subset $J = \{x_1,x_2\}$ parametrizes a singularity which needs to be resolved. According to Eq. \eqref{eq:dimregpartial-dimshift}, the $\lambda$-dependent monomials in the polynomials
\begin{align}
	{\widetilde{\mathcal{U}}}
	&= (\lambda x_1 + x_6)(\lambda x_2+x_3+x_4+x_5) + (\lambda x_2+x_4)(x_3+x_5)
	\quad\text{and}
	\label{eq:formfactor-U-scaled}\\
	{\widetilde{\mathcal{F}}}
	&= -s (x_1 x_4 x_5 + \lambda x_1 x_2 x_6 + x_1 x_3 x_6 + x_2 x_3 x_6 + x_1 x_4 x_6 + x_1 x_5 x_6)
	\label{eq:formfactor-F-scaled}
\end{align}
determine the distribution of dots in the dimension-shift identity for $F_{111111}(s, 4 - 2\epsilon)$ obtained by applying Eq. \eqref{eq:dimregpartial} to the integrand. Explicitly, we have
\begin{align}
	\label{eq:dimshift}%
	F_{111111}(s, 4 - 2\epsilon) 
	&=
-\frac{s}{2\epsilon}F_{221112}(s, 6 - 2\epsilon)
+ 3 F_{221111}(s, 6 - 2\epsilon)
	 \\ &\quad 
+ \frac{3}{2}F_{212111}(s, 6 - 2\epsilon)
+ \frac{3}{2}F_{122111}(s, 6 - 2\epsilon)
+ \frac{3}{2}F_{211211}(s, 6 - 2\epsilon) 
	\nonumber \\ &\quad 
+ \frac{3}{2}F_{211121}(s, 6 - 2\epsilon)
+ \frac{3}{2}F_{121121}(s, 6 - 2\epsilon)
+ \frac{3}{2}F_{121112}(s, 6 - 2\epsilon)\,.\nonumber
\end{align}
While the set $\{x_1,x_2\}$ is irrelevant for each of the integrals on the right-hand side of Eq. \eqref{eq:dimshift}, each of them still has subdivergences and further partial integrations are therefore required. 
In an optimized version of the {\tt HyperInt} setup, a singularity resolution of $F_{111111}(s, 4 - 2\epsilon)$ carried
out using the method of regularizing dimension shifts runs in a matter of seconds but returns an output which is more than twelve megabytes in size.
In fact, as we now demonstrate, exploiting IBP reductions leads to a drastically more compact result.

The first step is to use {\tt Reduze\;2} to reduce the integrals on the right-hand side of Eq. \eqref{eq:dimshift}.
According to our discussion in Section~\ref{sec:existence}, any of the integrals on the right-hand side of Eq. \eqref{eq:dimshift} could now be chosen as
a new master integral for the top-level sector; all eight integrals are guaranteed to be better behaved than $F_{111111}(s, 4 - 2\epsilon)$ with respect to the parameter subset $J$.
In the present example, however, we simply pick the corner integral in $6 - 2\epsilon$ dimensions, $F_{111111}(s, 6 - 2\epsilon)$, because it turns out to be a quasi-finite integral anyway and it is very convenient from
the point of view of IBP reduction.
After reduction, we obtain
\begin{align}
	\label{eq:dimshiftibp}%
	&F_{111111}(s, 4 - 2\epsilon) 
	=
	\frac{4(1-\epsilon)(3-4\epsilon)(1-4\epsilon)}{\epsilon s^2}F_{111111}(s, 6 - 2\epsilon) \\&\quad
	+ \frac{3(3-2\epsilon)(4-3\epsilon)(10-65\epsilon+131\epsilon^2-74\epsilon^3)}{\epsilon^2(1-2\epsilon) s^4} F_{101101}(s, 6 - 2\epsilon)\nonumber\\&\quad
	- \frac{6(3-2\epsilon)(5-3\epsilon)(4-3\epsilon)(14-119\epsilon+355\epsilon^2-420\epsilon^3+172\epsilon^4)}{(1-\epsilon)(1-2\epsilon)\epsilon^3 s^5} F_{100110}(s, 6 - 2\epsilon)\,.\nonumber
\end{align}
In Eq. \eqref{eq:dimshiftibp}, subsector integrals appear which have fewer propagators than the top-level integral.
In fact, as one can check by examining the relevant parameter subsets, $F_{100110}(s, 6 - 2\epsilon)$ is quasi-finite. $F_{101101}(s, 6 - 2\epsilon)$, on the other hand, is not.
In $6 - 2\epsilon$ dimensions, the regularizing dimension shifts produce an unusually simple, single-term result of the form
\begin{equation}
\label{eq:bubbletriangleres}
F_{101101}(s, 6 - 2\epsilon) = -\frac{12(2-\epsilon)(5-3\epsilon)}{\epsilon (1-\epsilon)} F_{103301}(s, 10 - 2\epsilon)\,.
\end{equation}

Finally, we can combine Eqs. (\ref{eq:dimshiftibp}) and (\ref{eq:bubbletriangleres}) to obtain an improved singularity resolution of the two-loop non-planar form factor integral,
\begin{align}
\label{eq:DandDres}
&F_{111111}(s, 4 - 2\epsilon) = \frac{4(1-\epsilon)(3-4\epsilon)(1-4\epsilon)}{\epsilon s^2}F_{111111}(s, 6 - 2\epsilon)  \\
&\quad - \frac{36(2-\epsilon)(3-2\epsilon)(4-3\epsilon)(5-3\epsilon)(10-65\epsilon+131\epsilon^2-74\epsilon^3)}{\epsilon^3(1-\epsilon)(1-2\epsilon) s^4} F_{103301}(s, 10 - 2\epsilon)\nonumber \\
&\quad- \frac{6(3-2\epsilon)(5-3\epsilon)(4-3\epsilon)(14-119\epsilon+355\epsilon^2-420\epsilon^3+172\epsilon^4)}{(1-\epsilon)(1-2\epsilon)\epsilon^3 s^5} F_{100110}(s, 6 - 2\epsilon)\,.\nonumber
\end{align}
Needless to say, a three-line singularity resolution is much better than one which is of order ten megabytes in size.
Eq. \eqref{eq:DandDres} clearly demonstrates the effectiveness of an IBP-based approach, but we shall soon see that it is actually not yet the minimal dimension shifts and dots resolution of interest.

The above construction uses regularity-improving dimension shifts explicitly.
As explained in the last section, this is not needed to arrive at a decomposition into quasi-finite integrals since, in general, it is a simple matter to determine suitable quasi-finite integrals by performing a direct search.
In fact, the systematic strategy described in Section \ref{sec:alg} reveals that  we could have chosen the simpler (with fewer dots and in smaller spacetime dimensions)
quasi-finite integrals $F_{102201}(s, 6 - 2 \epsilon)$ and $F_{100110}(s, 4 - 2\epsilon)$ to replace $F_{103301}(s, 10 - 2 \epsilon)$ and $F_{100110}(s, 6 - 2\epsilon)$.
The reduction in this new basis is readily computed to be, as we anticipated in \eqref{eq:formfactor-quasi-finite},
\begin{align}
\label{eq:DandDres2}
	F_{111111}(s, 4 - 2\epsilon) 
	&= \frac{4(1-\epsilon)(3-4\epsilon)(1-4\epsilon)}{\epsilon s^2}F_{111111}(s, 6 - 2\epsilon)  \\ &\quad
- \frac{10-65\epsilon+131\epsilon^2-74\epsilon^3}{\epsilon^3 s^2} F_{102201}(s, 6 - 2\epsilon)\nonumber \\ &\quad
- \frac{14-119\epsilon+355\epsilon^2-420\epsilon^3+172\epsilon^4}{(1-2\epsilon)\epsilon^3 s^3} F_{100110}(s, 4 - 2\epsilon)\nonumber
\end{align}
and is, by virtue of the fact that all three integrals on the right-hand side of Eq. \eqref{eq:DandDres2} were discovered by performing an exhaustive search,
guaranteed to be the minimal dimension shifts and dots resolution of the two-loop non-planar form factor integral (all other
quasi-finite integrals have either more dots or live in higher dimensions).
We checked \eqref{eq:DandDres2} with the exact results presented in reference~\cite{Gehrmann:2005pd}.

Taking into account the $\epsilon^{-1}$ singularities from the Gamma function prefactors,
we observe that, in both \eqref{eq:DandDres} and \eqref{eq:DandDres2}, the $\epsilon^{-4}$ and $\epsilon^{-3}$ poles of $F_{111111}(s, 4 - 2\epsilon)$ originate from subtopologies with at most four lines;
the six-line master integral $F_{111111}(s, 6 - 2\epsilon)$ enters the $\epsilon$-expansion only from the $\epsilon^{-2}$ pole onwards.
This suggests that a reduction to quasi-finite master integrals has a further benefit, namely that the computation of the $\mathcal{O}\left(\epsilon^i\right)$ term in the
$\epsilon$ expansion of a Feynman integral which has severe divergences in $4 - 2\epsilon$ dimensions may require fewer difficult parametric integral evaluations than in conventional 
approaches to multi-loop Feynman integral evaluation.

As a more involved example, let us consider the complete set of massless three-loop form factor integrals (their graphs are drawn, for example, in reference~\cite{1004.3653}).
By performing a systematic scan for simple quasi-finite integrals, which took only a few seconds, we find a minimal quasi-finite basis
of 22 master integrals with $0,\ldots,9$ dots in $d=4-2\epsilon,6-2\epsilon$ and $8-2\epsilon$ dimensions.
Particularly large numbers of dots happen to be required for factorizable topologies, which we treat as
three-loop integrals rather than considering their individual factors separately. 

\begin{figure}
\centering
\begin{gather*}
	b_1 = \Graph[0.43]{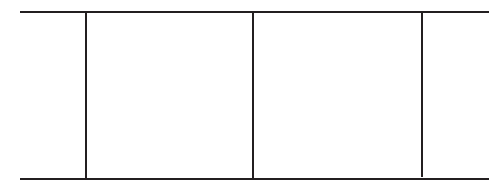}^{(6-2\epsilon)}
	\qquad
        b_2 = \Graph[0.43]{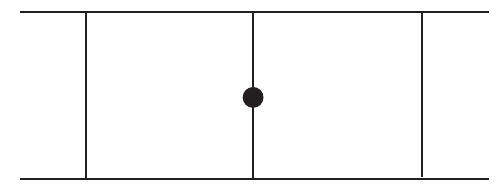}^{(6-2\epsilon)}
\end{gather*}
\begin{align*}
	b_3 &= \Graph[0.5]{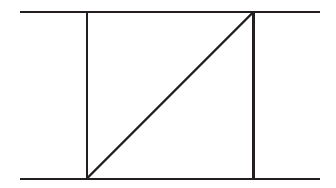}^{(6-2\epsilon)}
	&b_4 &= \Graph[0.5]{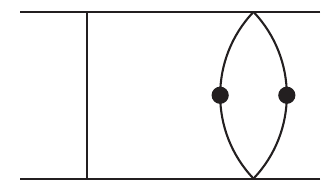}^{(6-2\epsilon)}
	&b_5 &= \Graph[0.45]{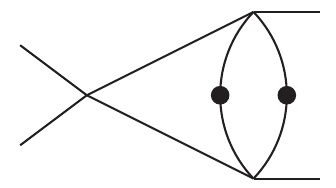}^{(6-2\epsilon)}
	\\[.65 cm] 
	b_6 &= \Graph[0.42]{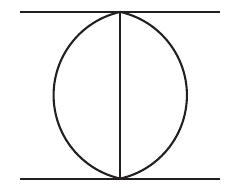}^{(4-2\epsilon)}
	&b_7 &= \Graph[0.42]{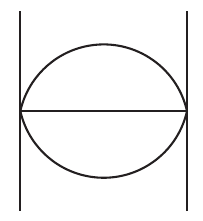}^{(4-2\epsilon)}
	&b_8 &= \Graph[0.42]{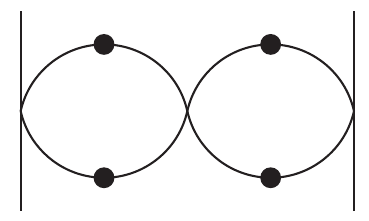}^{(6-2\epsilon)}
\end{align*}
\caption{A minimal quasi-finite basis for the planar massless double box integral family.
}
\label{fig:doubleboxmasters}%
\end{figure}

All examples considered so far had exactly one master integral per irreducible topology.
Our method works equally well if some sectors possess multiple master integrals or, in other words, several linearly independent integrals which live in the same topology.
For example, the two-loop planar massless double box top-level topology has two master integrals.\footnote{These integrals were first calculated in reference~\cite{Smirnov:1999gc} using the Mellin-Barnes method.}
Our method applies to the two-loop planar massless double box integral family because only two out of the three Mandelstam invariants, $s$, $t$ and $u$,
appear in the $\mathcal{F}$ polynomial of the top-level topology, which may be taken negative without violating the on-shell condition $s+t+u=0$.
This is enough for each term in the $\mathcal{F}$ polynomial to be positive definite in the interior of the integration region.
We find that it is straightforward to alter the basis of master integrals
given in reference \cite{SmirnovVeretin:DoubleBoxArbitrary} to obtain a minimal quasi-finite basis consisting of the eight Feynman integrals depicted in Figure \ref{fig:doubleboxmasters}.\footnote{%
The finiteness of the top-level master integrals ($b_1$ and $b_2$ in Figure~\ref{fig:doubleboxmasters}) was pointed out long ago in \cite{AnastasiouTauskTejedaYeomans:DoubleBoxIrreducible}.}
Let us emphasize that it makes a difference for the construction of a quasi-finite basis whether or not factorizable topologies are treated in the same way as all of the other topologies;
considering factorizable topologies as products of lower-loop topologies will typically lead to different results. In the present example, we treat the integral $b_8$ with the double bubble topology as a two-loop integral.
Considering each of its one-loop topologies separately would allow us to write it as the square of the (quasi-finite) one-loop bubble integral
in $4-2\epsilon$ dimensions without any dots at all. Finally, we remind the reader that, by virtue of the principle of analytical continuation, any quasi-finite linear combination equal to the original integral discovered in Euclidean kinematics 
can be reinterpreted as a relation between Feynman integrals in physical kinematics unambiguously using the $+i 0$ prescription.

\section{Outlook}
\label{sec:conclusions}

In this paper, we showed that one can express a multi-loop Euclidean Feynman
integral in terms of a basis of quasi-finite Feynman integrals.
Quasi-finite Feynman integrals, in the limit $\epsilon \to 0$, possess a convergent Feynman parameter representation 
except for a possible overall $1/\epsilon$ divergence encapsulated in a Gamma function prefactor.
These basis integrals are constructed for the original topology and its subtopologies
by allowing for higher spacetime dimensions and for higher powers of the propagators (dots).
Our new approach is guided by a regularization procedure~\cite{Panzer:2014gra}
introduced by the second author but, by employing integration by parts reductions, it overcomes practical limitations of the original method caused by runaway expression swell.
Our strategy, which we have dubbed the minimal dimension shifts and dots method,
is both efficient and straightforward to automate, and its implementation into
the public \texttt{Reduze\;2}~\cite{vonManteuffel:2012np} program is work in progress.

Our approach to singularity resolution can be viewed as an alternative to sector decomposition.
Crucially, in contrast to sector decomposition, our method cannot introduce spurious structures at an intermediate stage of a Feynman integral evaluation
because one never needs to split up the Feynman parameter representation of a quasi-finite integral.  
In particular, if the integral under consideration happens to be linearly reducible, one can subsequently apply powerful direct analytical integration techniques based
on the Feynman parameter representations of the quasi-finite integrals produced by the method~\cite{Panzer:2014caa}. In addition, it appears to be the case that sector decomposition produces many more convergent integrals
in typical situations than the minimal dimension shifts and dots method.

For completely general Feynman integrals it is not clear whether a quasi-finite
basis can be found with our algorithm, because physical kinematics could potentially introduce non-integrable divergences inside the integration domain.
In a heuristic approach, one could just follow the procedure and ultimately verify the quasi-finiteness of the candidates by trying to perform the integration over the Feynman parameters for $\epsilon=0$.
In any case, the IBP reduction and the decomposition into candidate integrals will be correct independent of whether or not the candidate integrals are actually quasi-finite.

For a wide class of multi-loop integrals, our decomposition may also be used for numerical evaluations.
The convergent parametric integrals generated by expanding in $\epsilon$ can, at least in principle,
be integrated numerically in the Euclidean case without further ado.
For integrals in physical kinematics, contour deformation~\cite{Borowka:2012yc} or other techniques may be employed,
provided the construction of a quasi-finite basis is possible along the lines discussed in this work.
Of course, further investigation is required, both to estimate the performance and stability of a numerical approach and to generalize it to Feynman integrals which do not admit a Euclidean region respecting the kinematical constraints.

\section*{Acknowledgements}
The research of RMS is supported in part by the ERC
Advanced Grant EFT4LHC of the European Research Council, the Cluster of Excellence Precision Physics, Fundamental Interactions and Structure of Matter (PRISMA-EXC 1098).
EP is supported by ERC grant 257638. Our figures were generated using {\tt Jaxodraw}~\cite{hep-ph/0309015}, based on {\tt AxoDraw}~\cite{CPHCB.83.45}.

\end{document}